%% file: main.tex
\documentclass[nonacm,sigconf]{acmart}
\usepackage{graphicx} 
\usepackage{listings}
\usepackage{xcolor}
\usepackage{ifthen}
\lstdefinestyle{customc}{
  belowcaptionskip=1\baselineskip,
  breaklines=true,
  frame=L,
  xleftmargin=\parindent,
  language=C,
  showstringspaces=false,
  basicstyle=\footnotesize\ttfamily,
  keywordstyle=\bfseries\color{green!40!black},
  commentstyle=\itshape\color{purple!40!black},
  identifierstyle=\color{blue},
  stringstyle=\color{orange},
}

\lstdefinestyle{customasm}{
  belowcaptionskip=1\baselineskip,
  frame=L,
  xleftmargin=\parindent,
  language=[x86masm]Assembler,
  basicstyle=\footnotesize\ttfamily,
  commentstyle=\itshape\color{purple!40!black},
}

\newcommand{\xnvme}{{\tt xNVMe}}

\newcommand{\iou}{{\tt io\_uring}}
\newcommand{\spdk}{{\tt SPDK}}
\newcommand{\review}{0}
\newcommand{\diff}[1]{
\ifthenelse{\review = 1}
{\textcolor{red}}
{}
{#1}
}

\title{\texttt{xNVMe}: Unleashing Storage Hardware-Software Co-design}
\author{Simon A. F. Lund}
\affiliation{
\institution{Samsung}
\city{Copenhagen}
\country{Denmark}
}
\email{simon.lund@samsung.com}
\author{Vivek Shah}
\affiliation{
\institution{Samsung}
\city{Copenhagen}
\country{Denmark}
}
\email{vi.shah@samsung.com}
\date{June 2023}
\setcopyright{none}
\begin{document}

\begin{abstract}
    NVMe SSD hardware has witnessed widespread deployment as commodity and enterprise hardware due to its high performance and rich feature set. Despite the open specifications of various NVMe protocols by the NVMe Express group and NVMe being touted as the new language of storage, there is a complex labyrinth of software abstractions to program the underlying hardware. The myriad storage I/O paths such as POSIX storage API, ad-hoc OS mechanisms, and userspace I/O libraries have different syntax and semantics that complicate software development and stand in the way of mass adoption and evolution of the NVMe ecosystem. 
    To unify the diverse I/O storage paths, we built \xnvme\ that exposes a single message-passing API to support both asynchronous and synchronous communication with NVMe devices. \xnvme\ provides various command sets to support diverse storage I/O paths in different OS (e.g., Linux, FreeBSD, Windows, and MacOS) and userspace libraries (e.g., \spdk) with minimal overhead. \xnvme\ is an Open Source project and has gained traction amongst various industry stakeholders. In this paper, we elaborate on the lessons that we have learned in the project during its evolution. We also provide some ongoing and future work planned for the project. We hope the database and storage systems community can join in the effort to both extend \xnvme\ and leverage it as a building block for innovative co-design of storage systems on modern NVMe hardware.
\end{abstract}

\maketitle
\pagestyle{plain}
\input{introduction}
\input{related-work}
\input{why-xnvme}
\input{what-is-xnvme}
\input{roadmap}

\section{Conclusion}
The diverse and fragmented storage I/O paths across different OS and userspace I/O libraries targeting NVMe hardware have created programmability challenges for storage systems. To unify the diverse storage I/O paths under a single, simple, and extensible API we have designed and built \xnvme. By utilizing a message-passing interface with support for various NVMe command sets, \xnvme\ allows flexible multiplexing of various storage paths across different OS and userspace libraries with minimal overhead. In this paper, we have presented various important design requirements of \xnvme\ that we have learned over time while engaging various storage I/O consumers. By simplifying the process of consumption and usage of various I/O paths, we hope to bring modern innovative NVMe hardware features closer to their consumption by storage systems.
\bibliographystyle{ACM-Reference-Format}
\bibliography{refs}
\end{document}

%% file: introduction.tex
\section{Introduction}
The past decade has witnessed the widespread evolution and deployment of NAND flash memory as commodity and enterprise storage hardware due to its high bandwidth and low latency. The growth of NAND flash SSDs has necessitated the birth of NVMe (Non-volatile memory express) access technology to sidestep the performance limitations of the SATA interface. The current global NVMe technology market share stands at 54.1 billion US\$ and is speculated to grow to 412 billion US\$ in 2031~\cite{nvme-market}. To coordinate the evolution and interoperability of NVMe technologies, the NVM Express working group was formed to create open specifications that could be implemented by hardware and software vendors~\cite{nvme-group}. Despite the open specifications of NVMe hardware, the challenge of software abstractions to program NVMe hardware remains and is growing with its popularity. 

Classically, the POSIX storage abstractions (\texttt{pread, pwrite}) had been the holy grail of programmability, stability, and portability to hide the underlying hardware complexity. However, the rise of diverse NVMe hardware has created a fissure in this perfect world. The need for a low-overhead, asynchronous programming model to leverage the high performance of modern NVMe hardware has created multiple complex I/O storage stacks differing from the original POSIX API. For the Linux kernel, there is the POSIX \texttt{aio\_*} APIs, \texttt{libaio}, and \iou\ storage interfaces\cite{io-uring:JoshiG0KRGLA24} that applications can program against. These interfaces differ widely in their API, semantics, and performance. The landscape of storage interfaces gets even more complicated if you factor in different OS.

The complexity of the storage I/O stack grows if you factor in support for newer SSD technologies such as ZNS SSDs, FDP SSDs, KV SSDs, and computational storage, to name a few. Either the rich-feature set is hidden behind the block layer interface of the OS or exposed via a userspace I/O stack such as \spdk, custom vendor libraries that are a thin shim over an NVMe device driver. This complexity and fragmentation of various storage I/O paths is unfortunate and creates unnecessary barriers in the adoption of the storage I/O paths by application stacks. The unnecessary complexity also stifles cooperation between software application designers and hardware vendors. \diff{As the usage of GPUs and hardware accelerators grows to support AI workloads, a similar API fragmentation is occurring for storage devices that require direct and efficient access by accelerators~\cite{smartio:MarkussenKHKSG21, bam:QureshiMGMMPXNV23}.}
 
\xnvme\ was envisaged to fill the programmability gap for NVMe storage technologies by creating a single unified API that applications can program against to flexibly multiplex the desired storage I/O path with minimal overhead~\cite{LundBJG22:storage-independence}. Instead of creating yet another abstraction for storage, \xnvme\ provides a single message passing API for interacting with NVMe devices along with support for various storage I/O paths using this API. Since storage systems are not locked into the API and semantics of a specific I/O path, they can flexibly experiment with different storage I/O paths based on need. \xnvme\ currently supports various I/O paths in the Linux kernel e.g., \texttt{libaio}, \iou\, and the classic POSIX abstractions of \texttt{pread} and \texttt{pwrite}. It also supports other OS e.g., FreeBSD and Windows and userspace I/O stack such as \spdk. \xnvme\ is an Open Source project~\cite{xnvme:github} that became publicly available in 2019. \diff{It began as an experimental platform for emerging NVMe interfaces (e.g., Open-Channel SSDs~\cite{lightnvm:BjorlingGB17}) and matured over time, gaining traction among academic researchers and industry practitioners.}

In this experience cum vision paper, we present our experiences and lessons learned during the project that has shaped its evolution. We outline the reasons behind its existence in Section \ref{sec:why} and some of the ongoing and future work in the project in Section \ref{sec:roadmap}. Our aim with \xnvme\ is to lower the entry barrier of building innovative data-intensive systems that leverage features of modern NVMe hardware. By providing I/O storage independence and an Open Source collaboration environment, we hope the project can foster co-design of data-intensive software systems and NVMe hardware than what is currently possible.

%% file: related-work.tex
\section{Related Work}
The initial concept and prototype of \xnvme\ was first presented to the Linux~\cite{vault} and SNIA ~\cite{xnvme_pesi_sdc} storage communities in 2020. \xnvme\ was later utilized as the experimental vehicle for the evolution of the Linux \iou\ command interface \cite{uring_cmd_sdc}. While the utility of \xnvme\ was acknowledged by the community, concerns were raised regarding the overhead of multiplexing various storage I/O paths. This was answered in our previous work~\cite{LundBJG22:storage-independence} where we have explored the architectural details and performance of \xnvme\ in more detail. In addition to the design, implementation and evaluation of \xnvme, our previous work of integrating \xnvme\ in upstream \spdk\ was presented to the \spdk\ community~\cite{bdevxnvme_spdkvf22}. In this paper, we have highlighted the reasons behind the design of \xnvme\ that we learnt by engaging with various consumers of I/O storage paths over the past few years. 

The ADIOS2 library~\cite{adios2} was devised to provide a uniform I/O framework to abstract different storage media (e.g., file, 
wide-area-network, in-memory staging, etc.) for supercomputer applications. Although both ADIOS2 and \xnvme\ aim to decouple applications from the underlying storage interface, ADIOS2 is specialized for scientific applications on supercomputers as it focuses on self-describing data variables and attributes. On the contrary, despite \xnvme\ being initially designed for NVMe hardware, it is not limited to NVMe devices and does not mandate any specific data format.

In the area of enabling emerging storage devices such as ZNS~\cite{nvme_zns}, libzbd~\cite{libzbd} and libnvme~\cite{libnvme} provide management operations via the Linux block and NVMe driver layers. Another option to enable ZNS would be to use a user-space driver such as \spdk. By contrast, \xnvme\ provides a single unified API for managing and accessing ZNS devices, over multiple system interfaces, including those utilized by the libraries described above.

The design and motivation of \xnvme\ has a lot of similarity with Intel's oneAPI~\cite{oneapi} programming model that was designed to provide a common interface over different accelerator devices. Similarly to \xnvme, oneAPI's design follows the hourglass model with Level Zero as its narrow waist. The main difference emnates from the abstractions they target. While \xnvme\ is designed for the I/O abstraction, oneAPI is designed for the compute abstraction. Nevertheless both their designs have similarities in device management, command contexts, command queues, and memory management. We think that these similarities are a confirmation that \xnvme\ is well designed to handle NVMe computational storage as well.

%% file: why-xnvme.tex
\section{Why xNVMe?}
\label{sec:why}
In this section, we provide the reasons that prompted the birth of \xnvme\ and shaped its evolution over time. These reasons are also important lessons that we have learned over time as we have tried to engage with various stakeholders in the hardware, OS, and application software communities to gauge the requirements of the project. 

\subsection{The Divergence of Storage Interfaces from POSIX}
\label{sec:why:posix-divergence}
The POSIX API was envisaged as the contract between application developers and the OS to write portable and efficient applications using the system call interface. The POSIX file-based abstractions for storage e.g., \texttt{pread,pwrite} have served for decades as the stable rock on which portable applications are built oblivious of the underlying storage technology. The block and character device abstractions and their APIs are minor variations of the file-based abstraction for storage in line with the UNIX philosophy of ``Everything is a file''. \textit{It is important to keep in mind that the POSIX API was developed in a different era of slow storage devices, limited storage hardware functionality, and limited applications}. We observe that over time significant major divergence has appeared among storage abstractions from the original POSIX API.

\subsubsection{The Challenge of Asynchronicity} The first major divergence appeared with the need for asynchronous abstractions for efficient I/O. Classically, the POSIX storage APIs had been designed to be synchronous where the calling code is blocked until the system call is completed. The need for efficient use of CPU cycles while waiting for I/O to complete raised the call for asynchronous abstractions. While initial attempts at asynchronicity used the \texttt{epoll/select} facilities to wait for interesting events on file descriptors before reading and writing, it only supported asynchronicity partially and the interface was unwieldy to use. This led to the POSIX async API proposals in the form of \texttt{aio\_*} system calls. Unfortunately, these APIs never really reached their intended goal of portability and widespread use.

FreeBSD supports POSIX aio while Windows supports an alternative asynchronous facility called IO control ports. Linux supports POSIX aio via a userspace glibc implementation that utilizes threads and has scalability and efficiency issues. Linux also provides \texttt{libaio} which is promoted as a programmable and efficient Linux-native alternative to POSIX aio. Of late, \texttt{io\_uring} has gained attention as a programmable asynchronous I/O abstraction in Linux for both userspace applications and within the kernel. \textit{If you are in search of a portable asynchronous abstraction for IO then POSIX has already failed you}.

\subsubsection{Fast Hardware and the Quest for High Performance} The second major divergence appeared with the emergence of fast hardware e.g., NVMe SSDs. Contemporary OS stacks had overheads in their design and implementation that prevented them from leveraging the raw hardware performance available. This gap was filled by the emergence of userspace storage I/O libraries notably SPDK and NVMe-Direct. Not only do these userspace I/O libraries provide a different API to the application but they also have significant implementation differences in how they interface and operate with the OS. SPDK has over time emerged as the storage IO library preferred by applications that require high storage performance but its API and design is a departure from the POSIX storage APIs. Recently, Linux has also taken note of the difference in performance offered by storage I/O libraries. This has accelerated the mainstreaming of \texttt{io\_uring} and NVMe I/O Passthru work in the kernel~\cite{io-uring:JoshiG0KRGLA24}. This line of work heavily borrows ideas from the userspace storage library stack e.g., shared-memory between userspace and kernel, asynchronous queuing mechanism for requests and responses, and polling to name a few. \textit{If you are looking to build fast portable storage services using modern high-performance hardware, POSIX has already failed you}.

\subsubsection{Emerging Feature-rich Storage Hardware}
The third major divergence appeared with the evolution of NAND Flash memory for storage e.g., NVMe SSDs. There are various types of emerging storage technologies e.g., ZNS SSDs, KV SSDs, Computational Storage SSDs that provide functionality beyond the traditional read-and-write abstraction of the block layer. These storage devices are designed for characteristics that can be leveraged by applications for efficient execution of application software code paths in hardware. ZNS SSDs are designed to remove the garbage collection overhead in the FTL so that well-behaved applications could reduce their write amplification and increase device lifetime. KV SSDs are designed to provide a key-value store implementation in hardware to reduce the overhead of software key-value store implementations. Computational storage hardware is designed to offload computations to the storage device from the host to reduce data movement and make efficient use of the computational capacity of storage devices. 

The designers of these storage technologies face the dilemma of exposing their functionality so that they can be leveraged by applications effectively. Wrapping them using a POSIX abstraction that was not designed for the use-case leads to the surrender of  all of their benefits. Moreover, proposing experimental extensions to well-established OS abstractions lead to resistance from OS maintainers who prefer stability over experimentation. Storage vendors are left with two choices to expose the functionality, namely (1) using a low-level NVMe device driver in an OS or (2) using a userspace I/O library e.g., SPDK. Both these choices limit the usability and reach of these features in the application space owing to programmability and deployment challenges. No matter how you expose the functionality, you have foregone the POSIX API. \textit{If you are looking to expose new storage hardware features to applications then POSIX has already failed you}.

\subsection{The Need for Unification of Storage I/O Paths}
\label{sec:why:unification}
In the previous section, we have provided a brief commentary on the major technological trends that have created myriad storage I/O paths with their API and semantics. \textit{This divergence from the unified POSIX API is unfortunate and has created a lot of portability and programmability problems}. Moreover, the presence of various I/O stacks each with its assumptions, syntax, and semantics introduce a steep learning cost and an architecture lock-in to the chosen storage I/O stack that bubbles up in the application. We need to maintain the original motivation of the POSIX API to create a unified API that can cater flexibly to the needs of all storage I/O paths. 

\subsubsection{Efficient Asynchronous API} The evolution of various storage interfaces in OS and userspace I/O stacks points to the centrality of a low-overhead asynchronous abstraction (API). There must also be support for a synchronous API to not impose the cost of
asynchronicity on applications that do not require it. While implementing a synchronous API using an asynchronous API might appear obvious to some, it is not trivial for a lot of users in practice. Careful API design can avoid this wasted effort. The API implementation must be efficient. The performance of the unified API should be the same as a specific I/O storage path API. Performance is important for adoption of the API by applications that want to seamlessly switch between the storage path of an OS and a userspace I/O library.

\subsubsection{Simple and Extensible} The presence of myriad storage I/O paths also points to the difficulty of conjuring a single future-proof storage abstraction with well-defined semantics and syntax. Instead, we believe the pragmatic approach would be to build a minimal API that allows extensibility. Such an approach can support various use-cases e.g., management of the devices vs reading/writing to them, and new experimental storage interfaces e.g., KV SSDs, Computational Storage devices, to name a few. The API should be agnostic to the specific storage I/O path and the underlying hardware technology so that applications can be portable and don't leak the underlying storage I/O path and hardware technology in their software architecture. 

\subsubsection{Flexible Multiplexing of I/O Paths} While the unification of storage I/O paths under a minimal API is worthy goal, it does not mean much if the API does not support various storage I/O paths that are currently used in practice. To make it easy to use, it should be simple and easy to switch the underlying storage I/O path to be used at run time.

It must be noted that we are not proposing an abstraction e.g., a block-device abstraction or a file abstraction with its bag of syntax and semantics. Instead, we are proposing a minimal API that allows applications to interact with various storage technologies using the desired storage path. The complexity of multiplexing of the storage I/O path should be hidden in the implementation of the API as much as possible than surfacing in the application. 

\subsection{The Many Faces of Storage I/O Consumers}
In the previous sections, we have highlighted the necessity of a simple, unified, and extensible API for various storage I/O paths. One of the central tenets in API design is to tailor the API for the consumers and their productivity needs. In this section, we highlight some of the diverse consumers of the storage I/O stacks and their use-cases. We observe that this focus on consumers is necessary for adoption and is a lesson that most software projects learn the hard way over time.

\subsubsection{Diverse Application Areas}
The first class of users are the traditional users of storage I/O paths i.e., the developers of various storage systems e.g., persistent key-value stores, database systems, userspace filesystems, and distributed filesystems, to name a few. These systems are classically built in C/C++ (also Rust of late) and have traditionally used OS abstractions for portability. Over time with the erosion of the POSIX interface support on different OS they have had to embrace the complexity of managing various ad-hoc storage I/O paths for their target deployments. Some of these systems would have also gained a lot from experimenting with diverse I/O storage stacks but the cost of changing I/O interfaces often acts as a deterrent for these application classes.

The second class of users are the developers of virtual machine monitors or type-2 hypervisors e.g., QEMU, Rust-VMM that facilitate the creation and operation of virtual machines on top of a physical host and are typically responsible for allocating and managing computing, memory and I/O devices on the physical host. These systems need to virtualize various storage I/O paths requested by the virtual machines. Instead of adding custom support for various storage paths and ensuring their correctness, these systems would benefit from having a single library and a unified API that allows them to flexibly change the desired storage I/O path. The need for the existence of such a library has often been articulated in their developer communities. QEMU is written in C while Rust-VMM is written in Rust.

The third class of users are storage hardware developers who propose and implement hardware specifications and testers who verify the implementation with the specifications. There are often teams with varying expertise e.g., hardware engineers who implement the spec in firmware (typically C) and testing teams who verify the specs often using scripting languages e.g., Python. These disparate teams have their own tooling experience and opinions on productivity. Prescribing a specific language or workflow often meets with resistance and loss of productivity.

The fourth class of users are the developers of standard and component libraries in various programming languages that provide high-performance storage APIs in the language by traditionally integrating with low-level storage I/O stacks e.g., OS system calls, userspace libraries etc. 

\textit{These various classes of users highlight the diverse languages and tooling through which the storage I/O stack is consumed. To ensure the least disruption amongst the users and the potential for high adoption, support for multiple programming languages is essential}.

\subsubsection{Idiomatic Programmable Storage API for Language Integration}
Our previous discussion has highlighted the need for a unified storage API to support multiple programming languages. This is critical for widespread adoption of the API because of the diverse ecosystem of users and tooling that consume the various storage I/O stack today. A simple and widely used model of exposing storage abstractions in OS and userspace libraries is to expose APIs in C and then use the language-specific facilities to integrate the C APIs. While such a solution provides simplicity to the implementers of the API, we note that such a solution can often be an impediment in adoption in today's diverse technological world. Most application programmers are comfortable with programming abstractions and practices in their language and find it ugly and unnecessary to understand the coding practices and assumptions of a foreign programming language e.g., C library APIs and the C programming language. These lead to costly misunderstandings regarding type transformations, memory management, and error handling during the integration of foreign language APIs which often doom the perception of the library/system being integrated.

It would be pragmatic for a storage API to provide bindings for multiple programming languages that can be directly consumed by its users. It is also possible to go a step further and ensure that the various language bindings of the storage API are "idiomatic" for the language they target. Such "idiomatic" bindings will conform to the widely accepted coding styles and practices of the language. The idiomatic nature of the API will ensure that consumers of the binding will have the experience of writing native programming language code and will be oblivious to the target language API that is encapsulated by the binding. An example would be to provide  language bindings in Rust that take care of the lifetime and manage the surface of \texttt{unsafe} annotations. Another example would be to provide Python bindings that conform to the widely used \texttt{with as} construct for variable management instead of explicit variable management. Another low-hanging fruit is to ensure typographical case compatibility (e.g., Camel case, Pascal case, Snake case, etc.) with the programming language. Such pragmatic (idiomatic) language bindings would reduce the barrier of entry of consumers of the storage API and facilitate its wide adoption.

\subsubsection{Low Barrier of Entry}
For any software project to gain widespread acceptability, it must have a low barrier of entry for new users and ensure ease of use. It is important that the software is easy to install and has a small list of dependencies. For many consumers of I/O storage stacks, the list of dependencies and the resultant impact on compilation times are extremely important since they have to operate in low-resource environments. One of the problems of new users of SPDK is its huge list of dependencies for all its supported functionality. While effort can be expended to prune the dependency list based on need, it involves a steep learning curve. Instead, it would be more user-friendly if the software provides a \textit{graceful degradation} experience where at compile time it lists the dependencies that are not present but makes progress with a minimal set of dependencies. During execution, it would signal an error if a dependency necessary to support the target storage path is not present. Such an approach would ensure that users only spend time for their particular use cases in a pay-as-you-go manner. Good documentation and developer support are also important for the success of the project. 

%% file: what-is-xnvme.tex
\section{What is xNVMe?}
\xnvme\ is a project which at the foundational level provides a C API and a cross-platform library implementing it. 
The API advocates a command-centric programming model designed around a message passing interface that is exposed through a queue abstraction to provide both asynchronous and synchronous API. 
On top of the \xnvme\ API are the NVMe command set specific APIs that help with command construction and packaging of NVMe semantics into the general \xnvme\ command infrastructure. These NVMe command sets are utilized in building command-line tools for interacting with NVMe devices as well as language bindings. 
In this section, we provide a brief exposition of the programming model of \xnvme\ for completeness. A curious reader can consult the \xnvme\ documentation~\cite{xnvme-docs} and our previous work~\cite{LundBJG22:storage-independence} which provides more details on the design, architecture, and implementation.

\subsection{Programming Model}
\begin{figure}
    \centering
    \includegraphics[width=0.5\linewidth]{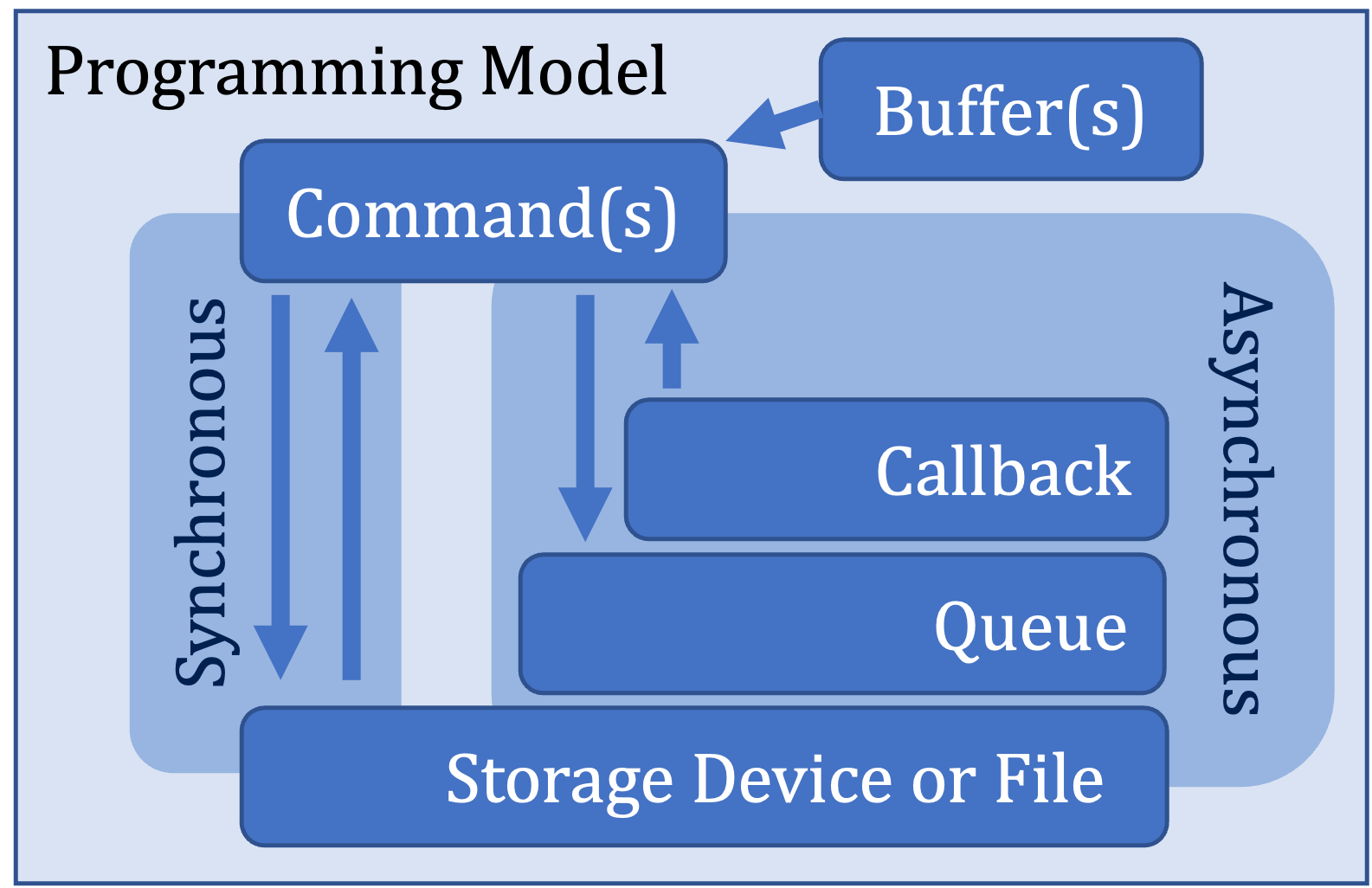}
    \caption{Components of the \xnvme\ programming model}
    \label{fig:enter-label}
\end{figure}

\xnvme\ revolves around expressions for shipping commands. To do so one needs: a device handle to send commands \textbf{to}, a buffer to carry the command-payload (if any). With these things in place, one can choose to send the command either in a synchronous manner "directly" to a device or in an asynchronous manner via a \textbf{queue}. The model is illustrated in Figure \ref{fig:enter-label}.

\subsection{C API Examples}
A full description of the \xnvme\ API is available on the \xnvme\ documentation page~\cite{xnvme-docs}. We provide some example usage of the API in a C-style pseudo-code to illustrate it. The example usage shows how to read a logical block address (lba) from an NVMe device.

A handle is represented as a \texttt{device} (\texttt{struct xnvme\_dev}) and obtained by providing a device identifier such as the operating-system device path (\texttt{/dev/nvme0n1}), a PCIe-address (\texttt{0000:03:00.0}), or a NVMe-transport endpoint (\texttt{10.11.12.42:4200}). In addition to device identifier, options are selected via \texttt{struct xnvme\_opts}.

\begin{lstlisting}[frame=single]
struct xnvme_opts opts = xnvme_opts_default();
struct xnvme_dev *dev;
dev = xnvme_dev_open("/dev/nvme0n1", &opts);
if (!dev)
    exit(errno);
\end{lstlisting}

The selection of I/O interface is done by \texttt{libxnvme} which decides based on the capabilities of the I/O storage paths available to the library at run-time. This default behavior is overridden by configuring options e.g. setting \verb|opts{.be.async = "io_uring"}|. In this case \texttt{libxnvme} will use \texttt{io\_uring} or fail in creating the device handle if the \iou\ library is not present. With a device-handle in place, one can start allocating payload buffers:

\begin{lstlisting}[frame=single]
struct xnvme_geo *geo = xnvme_dev_get_geo(dev);
size_t nbytes = geo->lba_nbytes;
void *buf;
buf = xnvme_buf_alloc(dev, nbytes);
if (!buf)
    exit(errno);
\end{lstlisting}

Here, a structure is retrieved from the device, \texttt{struct xnvme\_geo} that describes various characteristics of the device for the allocation of the payload buffer we are interested in the size of an \texttt{lba}. The buffer-allocator \texttt{xnvme\_buf\_alloc()}, provides a pointer to virtual memory, which is aligned according to the I/O constraints of the provided dev and any potential requirements of the I/O interface, such as DMA capability.

In the \xnvme\ C API a command-context (\texttt{struct xnvme\_cmd\_ctx}) encapsulates the \texttt{command} (\texttt{ctx.cmd}), the command-completion-status (\texttt{ctx.cpl}), the device to submit the \texttt{command} \textbf{to} (\texttt{ctx.dev}), or the queue to submit the \texttt{command} \textbf{via} (\text{ctx.queue}). The most significant command-field is the \texttt{opcode} (\texttt{ctx.cmd.opcode}) because it is a by-value encoding of the operation to be performed. The \texttt{command} is submitted in a synchronous manner. This is done by:

\begin{lstlisting}[frame=single]
struct xnvme_cmd_ctx ctx = {
    .dev = dev,
    .mode = XNVME_CMD_SYNC,
    .cmd.opcode = XNVME_SPEC_NVM_OPC_READ
};
err = xnvme_cmd_pass(&ctx, buf, nbytes, 0x0, 0);
if (err || xnvme_cmd_ctx_cpl_status(&ctx))
    exit(-err);
\end{lstlisting}

The error-handling required follows the \textbf{negative errno} return-value convention. In addition lower-level status-codes are embedded in the \texttt{ctx.cpl} which can be checked using the helper-function \texttt{xnvme\_cmd\_ctx\_cpl\_status()}.

To perform the operation in asynchronous fashion, a callback function and queue must be setup:

\begin{lstlisting}[frame=single]
static void callback(struct xnvme_cmd_ctx *ctx, void *cb_arg)
{
    if (xnvme_cmd_ctx_cpl_status(ctx))
        exit(EIO);
}

struct xnvme_queue *queue = NULL;
err = xnvme_queue_init(dev, 32, 0x0, &queue);
if (err)
    exit(-err);
xnvme_queue_set_cb(queue, callbackc, NULL);
\end{lstlisting}
Notice that in asynchronous mode, completion errors are handled in the callback function, and submission errors are handled by the return-value of \texttt{xnvme\_cmd\_pass()} as done here:

\begin{lstlisting}[frame=single]
struct xnvme_cmd_ctx *ctx;
ctx = xnvme_queue_get_ctx(queue);
ctx->cmd.opcode = XNVME_SPEC_NVM_OPC_READ;
submit:
err = xnvme_cmd_pass(ctx, buf, nbytes, 0x0, 0);
switch(err) {
case 0:
    break;
case -EBUSY:
case -EAGAIN:
    goto submit;
default:
    exit(-err);
}
\end{lstlisting}

The above examples briefly illustrate the \xnvme\ programming model. They also show the complexity introduced by changing blocking semantics into non-blocking. Note that this is not a deficiency introduced by \xnvme\, rather, this is an intrinsic characteristic of non-blocking APIs as the submission and completion stage are separated. The API unification and reduction in complexity is provided here underneath the \xnvme\ \texttt{queue} where lies implementations of the myriad of asynchronous queue-representations and APIs.


\subsection{I/O Storage Path Support}
\xnvme\ currently supports a wide array of I/O storage paths targeting OS such as POSIX aio, \texttt{libaio}, Windows IOCP, Windows IORING, \iou, \texttt{io\_uring} command, psync (\texttt{pread,pwrite}), block-layer IOCTLs, NVMe driver layer IOCTLs. These storage paths are supported on compatible OS such as Linux, FreeBSD, Windows, and MacOS. To add to the OS managed storage paths are the userspace NVMe drivers from \spdk\ and \texttt{libvfn}. The same unchanged C API examples outlined previously will work with all the storage I/O paths outlined above. In case the API is invoked on a non-NVMe device, then a fallback shim layer maps the NVMe commands to appropriate OS managed I/O operations.  

\subsection{Language Support}
\xnvme\ is implemented in C for its low-level of control for system programming and its rich mechanics for language-interoperability via C-ABI compatible foreign function interfaces. Currently, C++ can directly consume the \xnvme C API. Additionally, bindings are provided for Python and work is planned to expose bindings in Rust. We are also looking to make the bindings idiomatic to reduce the impedance in integrating the bindings in different programming languages.


%% file: roadmap.tex
\section{The Road Ahead for \protect\xnvme}
\label{sec:roadmap}
In this section, we outline some of the interesting avenues of work in the project that might be interesting to the research community. By outlining the project roadmap, we hope to engage various stakeholders in future directions of the project.

\subsection{Efficient Storage Access from Accelerators}
\diff{
Efficient and fast storage access from GPUs has recently gained attention as AI workloads continue to explode. This trend has driven demand for direct, low-cost access to storage with minimal CPU and main memory overhead. However, recent efforts~\cite{bam:QureshiMGMMPXNV23, smartio:MarkussenKHKSG21} have produced ad hoc programming models and APIs that not only require familiarity with low-level hardware interfaces, such as PCIe and NVMe, but also tightly couple applications to specific hardware technologies.}

\diff{
While these solutions leverage advanced hardware capabilities effectively, their diversity echoes the fragmentation seen with POSIX storage APIs following the introduction of newer NVMe SSDs. Based on our experience with \xnvme\, we view this trend as an opportunity for xNVMe to establish a standardized programming model and API for GPU applications—one that manages underlying hardware seamlessly while enabling efficient, direct storage access. We are actively exploring abstractions and implementations that facilitate data exchange between GPUs and modern NVMe storage devices.}

\subsection{Idiomatic Language Bindings}
Currently, \xnvme\ supports an asynchronous storage API in C. Our experiences with stakeholders of various software projects integrating the C API in various programming languages has highlighted the potential for idiomatic language bindings. We are exploring the generation of idiomatic language bindings for languages like C++, Rust, Java, and Python. While pursuing this endeavor we are developing a binding generator tool~\cite{yace:github} to ease the process of generation and maintenance of language-specific bindings. 

\subsection{Integration with Open Source Data Management Systems}
Data management systems have traditionally been the most sophisticated consumers of high-performance and feature-rich hardware. Applications rely on the stability and performance of data management systems for their daily work. We want to explore the potential of integrating \xnvme\ in popular Open Source data management systems. Usage of \xnvme\ will allow the systems to flexibly utilize the various encapsulated storage I/O paths based on need. Since \xnvme\ does not require an architectural lock-in and supports both POSIX I/O and userspace libraries, it can serve as a good starting point to abstract storage endpoints in data management systems targetting NVMe hardware. Some initial work has been done in this direction to integrate RocksDB with xNVMe for conventional and ZNS SSDs and we are looking to integrate more systems.

\subsection{Virtual Machine Monitor Integration}
There has been interest amongst developers of virtual machine monitors e.g., QEMU, rust-vmm~\cite{rust-vmm:github} to use a library to emulate various storage I/O paths instead of implementing them in various virtual machine monitor software implementations. Initial work has been done on this front in the libblkio library~\cite{libblkio:gitlab} which has used \xnvme\ as the underlying library to encapsulate various storage I/O paths. There is an open pull request in \texttt{libblkio} for this integration effort and more work is planned to evaluate this integration.

\subsection{Computational Storage Support}
Computational storage devices have gained attention as a technology to leverage the computational capacity of the storage devices by offloading computation from the host device to the storage device. However, these technologies are quite niche from an application perspective and require an understanding of complex low-level and fragile software stacks for applications to meaningfully use them. Work is ongoing in \xnvme\ to bring support for computational storage devices meeting the NVMe Computational Storage specification. Most of the work has already been done and will be released publicly when the specification is ratified by the NVMe board.